# Oxygen vacancy engineering in pulsed laser deposited BaSnO$_3$ thin films on SrTiO$_3$

Wilson Román Acevedo[1], Myriam H. Aguirre[2,3,4], Beatriz Noheda [5,6], Diego Rubi[1,*]

[1]*Instituto de Nanociencia y Nanotecnología (INN),*

*CONICET-CNEA, Gral. Paz 1499, 1650 San Martín, Argentina*

[2] *Instituto de Nanociencia y Materiales de Aragón (INMA-CSIC),*

*Campus Rio Ebro C/Mariano Esquillor s/n, 50018 Zaragoza, Spain*

[3] *Dpto. de Física de la Materia Condensada,*

*Universidad de Zaragoza, Pedro Cerbuna 12, 50009 Zaragoza, Spain*

[4] *Laboratorio de Microscopías Avanzadas, Edificio I+D,*

*Campus Rio Ebro C/Mariano Esquillor s/n, 50018 Zaragoza, Spain*

[5] *CogniGron - Groningen Cognitive Systems and Materials Center,*

*University of Groningen, Nijenborgh 4,*

*9747AG Groningen, The Netherlands*

[6] *Zernike Institute for Advanced Materials,*

*University of Groningen, Nijenborgh 4,*

*9747AG Groningen, The Netherlands* *

## Abstract

We demonstrate the tunability of oxygen content in pulsed laser deposition (PLD)-grown barium stannate (BaSn$O_3$, BSO) thin films by precisely controlling the background oxygen pressure over a broad range from 0.0004 mbar to 0.13 mbar. The introduction of oxygen vacancies significantly alters the structural properties of BSO films, inducing a monotonic expansion of the out-of-plane lattice parameter and cell volume as the vacancy concentration increases. The progressive formation of oxygen vacancies was spectroscopically tracked using X-ray photoelectron spectroscopy (XPS), providing direct insight into the vacancy evolution. Furthermore, we show that the oxygen stoichiometry in BSO plays a critical role in modulating the sheet resistance of BSO/LaScO$_3$ heterostructures, enabling interface metallic electron conduction. This oxygen content control offers a robust strategy to tailor the electronic properties at the interface, highlighting its potential for oxide electronics and functional interface engineering.

BaSn$O_3$ (BSO) is an oxide perovskite that has gained significant interest due to its exceptional electronic and optical properties, such as high room temperature carrier mobility (values up to 320 cm$^2$/Vs have been reported after lanthanum doping [1], significantly larger than the ≈ 10 cm$^2$/Vs typical of other perovskites such as SrTiO$_3$ [2]) and large optical bandgap (≈ 3.1 eV [3]), making it transparent in the visible region. These properties turn BSO as highly suitable for the next-generation of optoelectronic devices, transparent conducting oxides and high-mobility oxide transistors [4].

BSO presents a cubic structure with cell parameter a = 4.116 $\vec{A}$ and is chemically and thermally stable in air, making it ideal for high-temperature and harsh-environment applications. BSO has been synthesized in thin film form by using different techniques including molecular beam epitaxy [5–7], pulsed laser deposition [8–11] and sputtering [12, 13], either in single layers or integrated with other perovskite oxides in heterostructures, leading, for example, to the formation of 2-dimensional electron gases (2DEG) at oxide interfaces with LaAlO$_3$ [14] and LaScO$_3$ [15]. These systems display room temperature mobilities of 18 cm$^2$/Vs and 60 cm$^2$/Vs, respectively, higher than the reported values for the canonical oxide-based 2DEG SrTiO$_3$/LaAlO$_3$ system (≤ 10 cm$^2$/Vs at room temperature) [16, 17]. This highlights the high technological interest that present 2DEG systems involving BSO. The 2DEG stabilization mechanism on these systems was attributed either to an interface electronic reconstruction -triggered by the existence of the so called "polar catastrophe"- [15] or to the presence of oxygen vacancies (OV) at the interface [14], which inject electrons and can substantially increase the (2D) carrier density [14]. These two effects might compete, resembling the case of other oxide-based 2DEG systems such as $SrTiO_3/LaTiO_3$ [18] or $SrTiO_3/LaAlO_3$, which presents a complex phase diagram for different carrier densities that also includes superconductivity and magnetism [19].

* diego.rubi@gmail.com

OV are ubiquitously found in oxides [20], particularly in thin films, and play a fundamental role in, for example, applications such as resistive memories and memristors, both based on the resistive switching effect [21, 22], or in magnetic thin films where they can tune the ferromagnetic interactions (i.e. by shifting the Curie temperature [23]). In the case of BSO, the presence of OV can be used as a knob to tune the bandgap [24] -and eventually trigger sub-bandgap absorption- or to modify both the electron mobility and electrical conductivity [13, 25], indicating that proper OV engineering can lead to the optimization of different functional properties of BSO-based devices. This requires both a careful control of the fabrication procedure in order to warrant sample repeatability and a thorough understanding of the effect of the OV on the materials physical and chemical properties.

In this paper, from a combination of experimental techniques including x-ray diffraction, x-ray photoemission spectroscopy and transmission electron microscopy we show how the controlled introduction of OV on BSO epitaxial thin films grown on $SrTiO_3$ at different oxygen pressures impacts on the films' structural, spectroscopic and transport properties. We show that, altougn BSO single layers remains electrically insulating in all cases, OV engineering can be used to tailor the room temperature sheet resistance of 2DEG occurring at the $BSO/LaScO_3$ interface.

BSO thin films were grown on $SrTiO_3$ (STO) single crystals by Pulsed Laser Deposition (PLD) assisted by Reflection High-Energy Electron Diffraction (RHEED). The deposition temperature was set in 760 °C and the oxygen background pressure was varied between 0.13 and 0.0004 mbar. Figure 1(a) displays an x-ray reflectivity experiment performed on a sample grown at 0.04 mbar of $O_2$. From the fitting of these measurements, the thickness of the sample series grown at different oxygen pressures was estimated to be in the range 16-23 nm. Figure 1(b) displays an atomic force microscopy topography of the sample grown at 0.04 mbar, where a grainy structure with root mean squared roughness (RMS) of 1.1 nm is appreciated. The RMS roughness corresponding to the other samples of the series was in the range 0.9-1.3 nm. Figure 1(c) shows a x-ray diffraction pattern in Bragg-Brentano configuration, corresponding to the 0.04 mbar sample, with the only presence of (00l) peaks coming from both the substrate and the BSO layer, reflecting the epitaxial nature of the latter. Finally, Figure 1(d) displays the RHEED pattern of the same film, recorded in-situ after growth, showing a spotty structure consistent with the low roughness extracted from the atomic force microscopy image of Figure 1(b).

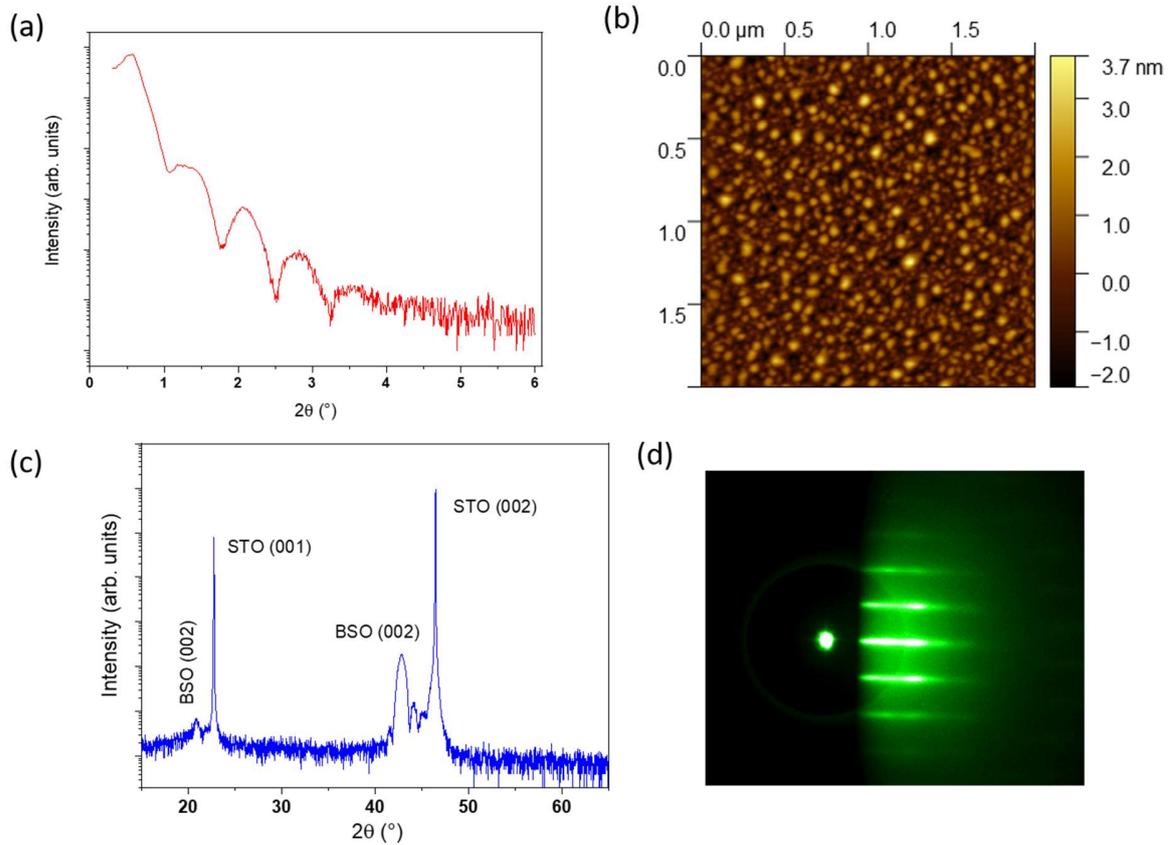

FIG. 1: (a) X-ray reflectivity experiment corresponding to a BSO thin film grown at 0.04 mbar of oxygen; (b) Atomic force microscopy topography and (c) X-ray diffraction pattern in Bragg-Brentano geometry corresponding the the same sample; (d) RHEED pattern taken in-situ after the growth of the same sample. A spotty pattern is seen, indicating the presence of low surface roughness.

Figure 2(a) shows blow-ups of the x-ray specular diffraction patterns recorded on the BSO film grown in the entire range of explored oxygen pressures, showing the evolution of the (002) peaks corresponding to both the STO substrate and the BSO thin layer. It is seen that as the growth oxygen pressure decreases, the BSO (002) peaks shift to lower angles, reflecting an expansion of the out-of-plane cell parameter. In order to track the evolution of the in-plane cell parameter, we have measured Reciprocal Space Maps (RSM) around the (103) non-specular reflections of both STO and BSO. Figure 2(b) shows an example of the latter, corresponding to the BSO film grown at 0.08 mbar. The extracted out-of-plane and in-plane cell parameter are displayed in Figure 2(c). It is found that the in-plane remains roughly constant around 4.11 Å, in good agreement with the bulk cell parameter. This indicates that the BSO unit cell is relaxed in-plane, as expected from the large mismatch existing between STO (with cell parameter a = 3.906 Å) and BSO cubic structures (≈ 5.8 %), which prevents the films from being fully strained. It has been shown that these strain relaxations occur via the formation of extended defects such as threading dislocations [6, 26, 27]. As mentioned before, the BSO out-of-plane cell parameter expands from 4.16 Å for the films grown at 0.13 mbar, to 4.23 Å for the films grown at 0.0004 mbar, indicating the progressive formation of oxygen vacancies which induce a unit cell enlargement, as a consequence of the increased electrostatic repulsion between cations. Additionally, these values are larger than the BSO bulk cell parameter in all cases, indicating the presence of compressive in-plane strain, as expected from the positive mismatch existing between BSO and STO. Figure 2(d) shows a STEM high resolution cross-section of the interface between STO and BSO, with the latter grown at 0.08 mbar. A FEI Titan G2 microscope with probe corrector (60-300 keV) was used. The epitaxial nature of the BSO layer is confirmed by the Fast Fourier Transforms displayed as insets. Additionally, on the right side of the image it is seen the presence of an extended defect (seen with light contrast) in the BSO layer, related to the already mentioned large mismatch between STO and BSO and to the release of elastic energy at the BSO layer.

In order to confirm the controlled incorporation of oxygen vacancies in the BSO structure, we have performed x-ray photoemission spectroscopy (XPS) on the BSO films grown at different oxygen pressures. The XPS measurements were performed on a homebuilt system equipped with a VG Microtech CLAM 2 hemispherical analyzer using a non-monochromatic Al K$\alpha$ source (1486.6 eV, operating at 10 kV, 34 mA, 30° source). In all cases, Shirley or Shirley/Linear functions were used to model the background. Binding energy shifts were corrected by using the position of the adventitious C 1s peak (284.8 eV) [28]. Figures 3(a) and (b) display O 1s spectra recorded on samples grown at 0.04 and 0.0004 mbar, respectively. The spectra display an asymmetric shape that can be modeled by assuming the presence of two Gaussian-Lorentzian (50%/50%) peaks; a lower binding energy one related to regular lattice oxygen and a higher binding energy one associated to oxygen atoms in the vicinity of oxygen vacancies [24, 25, 29, 30]. To ensure that the fittings remain physically meaningful, we have fixed the distance between the position of both peaks at 1.4 eV, and constrained their full-width-hall maxima to differ in less than 0.2 eV [31]. As the growth oxygen pressure decreases, the vacancy-related peak grows, reflecting the increasing amount of oxygen vacancies incorporated in the BSO layer. This is confirmed by Figure 3(c), which plots the evolution of the area of the vacancy-related peak with respect the total O 1s peak as a function of the growth oxygen pressure. Figure 3(d) shows, in addition, that chemical shifts towards lower binding energy occur for both regular and vacancy-related oxygen peaks as the growth oxygen pressure lowers. When oxygen vacancies are created, the material undergoes lattice expansion (recall the structural data previously discussed), which leads to longer metal-O bond lengths and, therefore, a reduction of the the covalent character of the bonds is expected. As the bonding becomes more ionic, the oxygen atoms experience a lower effective nuclear charge from the surrounding cations, as the electron cloud becomes more spread out, resulting in lower binding energy for the O 1s electrons.

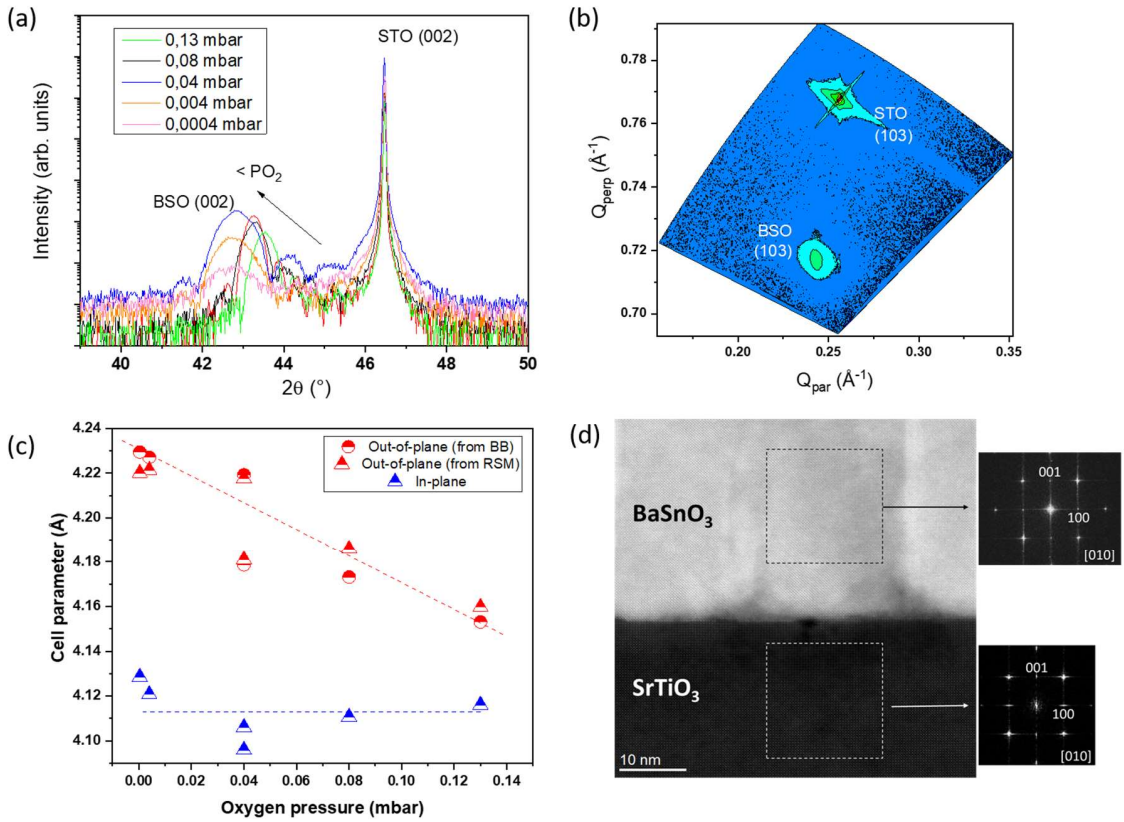

FIG. 2: (a) X-ray diffraction patterns in Bragg-Brentano configuration corresponding to BSO samples grown at oxygen pressures between 0.0004 and 0.13 mbar. The presence of (002) reflections arising both from the STO substrate and BSO films are seen, consistently with the epitaxial nature of the structures; (b) Reciprocal Space Maps recorder around the (103) non-specular reflections of both STO and BSO, with the latter grown at an oxygen pressure of 0.08 mbar; (c) Evolution of in-plane and out-of-plane cell parameters for BSO films as a function of the growth oxygen pressures. Data was extracted from both x-ray Bragg-Brentano and Reciprocal Space Maps experiments; (d) High resolution STEM cross section corresponding to a BSO films grown on STO at 0.08 mbar. The blow-ups show Fast Fourier transform performed on both the film and the substrate, confirming the epitaxial growth of BSO.

Figures 3(e) and (f) display the XPS spectra corresponding to the Ba 3d and Sn 3d deep levels for the BSO sample grown at 0.04 mbar. In both cases, the existence of single doublets, originated in spin-orbit splittings of 15 and 8.4 eV, respectively, is seen. The same results were obtained for samples grown at other oxygen pressures. The binding energy of these peaks did not display any significant variations between different samples. The recorded data suggest that the excess of electrons injected by oxygen vacancies may increase the lattice's electron density without changing the formal valence of Sn (+4). Moreover, the absence of a chemical shift in the Sn 3d levels upon oxygen vacancy creation and lattice expansion further evidences the stability of the valence for this cation, and suggests that changes due to vacancies primarily affect the oxygen sublattice.

In summary, the controlled and progressive oxygen vacancy incorporation to the BSO lattice was infered from both structural and spectroscopic data. Interestingly, room temperature van-der-Pauw measurements performed in the whole series of BSO samples shows that the oxide maintains a highly insulating behavior, with sheet resistance above 1 $G\Omega$/sq, for all the explored oxygen pressures. This is in contrast with Ref. [14], which shows the existence of windows of oxygen growth pressures (between 0.0004 and 0.003 Torr) where BSO resistivity largely drops. It is argued that in this pressure window oxygen vacancies might act as effective shallow donors.

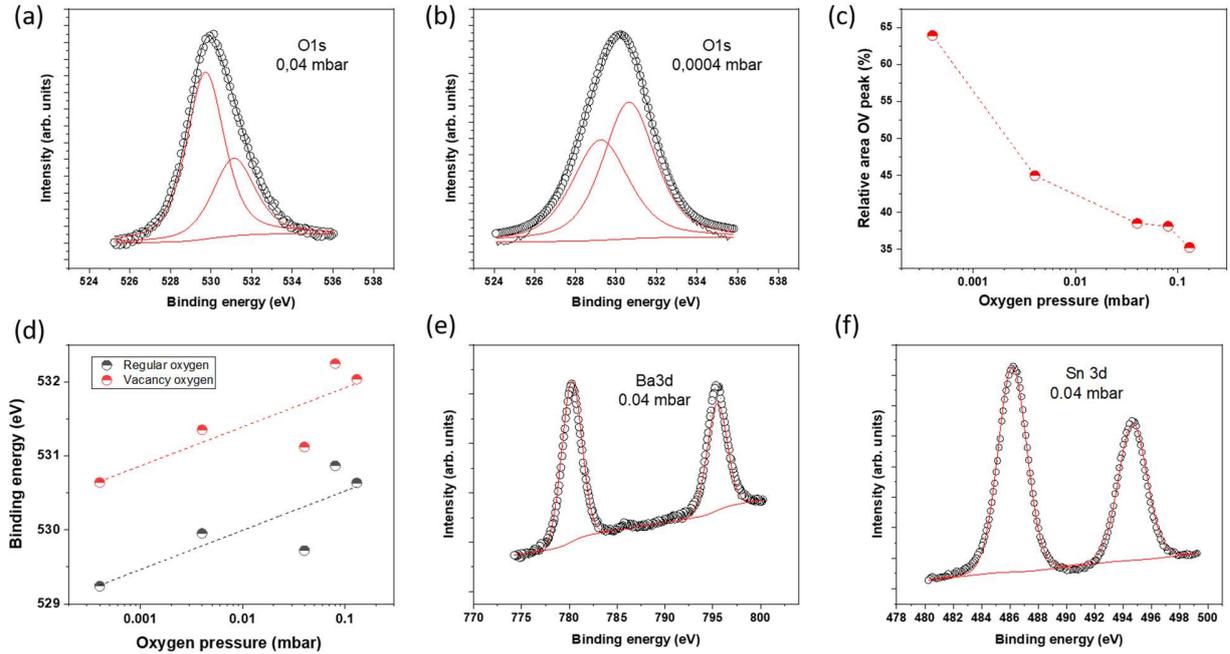

FIG. 3: (a), (b) X-ray photoemission spectra corresponding to O 1s level for BSO thin films grown at 0.04 and 0.0004 mbar, respectively. The spectra was simulated by convoluting two peaks, shown in red lines, one corresponding to regular lattice oxygens and the other to oxygen ions close to oxygen vacancies; (c) Evolution of the relative area of the latter peak as a function of the growth oxygen pressure. The progressive creation of oxygen vacancies as the oxygen pressure decreases is evidenced; (d) Evolution of the binding energy of regular and oxygen vacancy peaks as a function of the growth oxygen pressure. A chemical shift towards higher binding energies for higher growth oxygen pressures is seen in both cases; (e), (f) X-ray photoemission Ba 3d and Sn 3d measurements corresponding to a BSO thin films grown at 0.04 mbar. In both cases, the spectra can be fitted by assuming a single doublet.

In order to test the effect of BSO oxygen stoichiometry on the electronic properties of a BSO-related 2DEG, we have grown by laser ablation STO/BSO/LaScO$_3$ (LSO) multilayers, which were recently shown to display interface metallic conductivity, attributed to the existence of a polar catastrophe, in heterostructures grown by molecular beam epitaxy [15]. Both layers were grown at 760 °C, while the BSO oxygen pressure was varied between 0.13 and 0.04 mbar, maintaining the LSO growth pressure constant at 0.1 mbar. The BSO and LSO thicknesses were ≈ 300 and 10 nm, respetively. Figure 4(a) displays a Bragg-Brentano x-ray diffraction scan of one of the fabricated heterostructures, displaying only (00h) reflections for both BSO and LSO layers, reflecting the epitaxial nature of the bilayer. The inset shows a HAADF-STEM cross section where the presence of a well defined, sharp interface between both perovskites is seen. Figure 4(b) displays the evolution of the van-der-Pauw [32] room temperature sheet resistance, displaying a several orders of magnitude drop for the heterostructures, consistent with the formation of a 2DEG at the BSO/LSO interface. Noticeably, the sheet resistance decreases as the BSO growth pressure decreases (from ≈ 12.5 kΩ/sq for BSO grown at 0.13 mbar to ≈ 150 Ω/sq for BSO grown at 0.01 mbar), suggesting that the interface conductivity is strongly determined by the presence of oxygen vacancies and the electron carriers left behind by missing oxygen ions, increasing the carrier density. This effect is likely intertwined with the one arising from the polar catastrophe previously reported by Eom et al. [15]. The presence of oxygen vacancies is also expected to affect the electron mobility, as the probability of carrier scattering processes at point defects might increase. A comprehensive electrical characterization, including the temperature dependence of sheet resistance, carrier density, and mobility, along with a high-resolution TEM microscopy study of other critical effects such as cation intermixing at the interface [33], will be reported elsewhere.

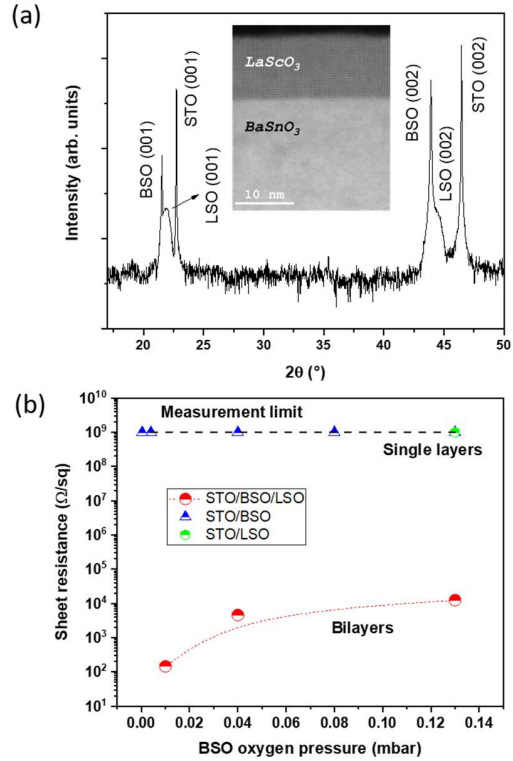

FIG. 4: (a) Bragg-Brentano x-ray diffraction pattern corresponding to a STO/BSO/LSO bilayert. The inset shows a HAADF-STEM cross-section corresponding to the same structure; (b) Evolution of the room temperature sheet resistance, measured in Van der Pauw configuration, of BSO/LSO interfaces for different BSO growth oxygen pressures. The figure also displays the values measured on both BSO -grown at different oxygen pressures- and LSO single layers, which show a highly insulating behavior.

In summary, we demonstrated that the oxygen content of PLD-grown BSO thin films can be effectively tuned by controlling the background oxygen pressure over a wide range from 0.0004 mbar to 0.13 mbar. The introduction of oxygen vacancies significantly influences the structural properties of the films, leading to a monotonic expansion of the out-of-plane lattice parameter and cell volume as the vacancy concentration increases. The progressive formation of oxygen vacancies was also monitored spectroscopically through XPS experiments. Lastly, we showed that the oxygen content in BSO critically governs the sheet resistance of BSO/LSO heterostructures, enabling interface metallic electron conduction. This tunability offers a powerful approach to control the electronic properties of the interface, including carrier density and mobility.

We acknowledge support from ANPCyT PICT2019-02781 and EU-H2020-RISE project MELON (Grant No. 872631). The financial support from the Groningen Cognitive Systems and Materials Center (CogniGron) and the Ubbo Emmius Foundation of the University of Groningen is appreciated. We thank M. Salverda for the access to the XPS facility at the University of Groningen.

---